\documentstyle[sprocl]{article}
\input{psfig}
\def\Journal#1#2#3#4{{#1} {\bf #2}, #3 (#4)}

\def\NPB{{\em Nucl. Phys.} B}

\def\PRD{{\em Phys. Rev.} D}
\def\ZPC{{\em Z. Phys.} C}

\def\be{\begin{equation}}
\def\ee{\end{equation}}
\def\bea{\begin{eqnarray}}
\def\eea{\end{eqnarray}}

\begin{document}
\title{ELECTROPRODUCTION OF HEAVY QUARKS AT NLO$\;$\footnote{Presented at 
the 1996 Meeting of the American Physical Society, Division of Particles 
and Fields (DPF 96), Minneapolis, Minnesota, 10-15 Aug 1996.}}
\author{BRIAN W. HARRIS}
\address{Physics Department, Florida State University, \\ Tallahassee,
FL 32306-3016, USA}
\maketitle

\abstracts{A new next-to-leading order Monte Carlo program for the 
calculation of fully differential heavy quark cross sections in 
electroproduction is described.  A comparison between the theoretical 
predictions and the latest charm production data from H1 and ZEUS at 
HERA is presented.}

\section{Introduction}
Heavy quark production in deeply inelastic electron proton scattering 
(DIS) has long been recognized as a candidate process for extracting 
the gluon distribution in the proton because the process is almost 
exclusively dominated by photon-gluon fusion \cite{emc}.  The cross 
section for heavy quark production in DIS is given in terms of the 
heavy quark structure functions $F_2(x,Q^2,m^2)$ and $F_L(x,Q^2,m^2)$.
A calculation of these in next-to-leading order (NLO) QCD has been
available for several years \cite{lrsvn}.  However, only recently have
they been calculated in a fully differential form at NLO \cite{hs}.  
The relation between the cross section and the structure functions, in 
terms of the usual DIS kinematic variables $x$, $y$ and $Q^2$, is as follows:
\be
\label{eq:cross}
\frac{d^2\sigma}{dxdy} 
= \frac{2\pi\alpha^2}{Q^4} S \left[ \left\{ 1 + (1-y)^2 \right\} 
F_2(x,Q^2,m^2) - y^2 F_L(x,Q^2,m^2) \right] \, .
\ee
The electromagnetic coupling is $\alpha = e^2 / 4 \pi$,  
$\sqrt{S}=300\; {\rm GeV}$ at HERA, and $m$ is the heavy quark mass.

\section{Method}
The results presented herein are based on the fully differential structure 
functions \cite{hs} and Eq.~\ref{eq:cross}.  The computation was performed 
using the subtraction method.  Therefore, all phase space integrations 
may be evaluated using standard Monte Carlo techniques to 
produce histograms for fully differential, semi-inclusive, or inclusive 
quantities related to any of the outgoing particles.  Experimental cuts 
can also be imposed.  When considering the production of charmed mesons, 
fragmentation effects are modeled using the formalism of Peterson 
{\it et al} \cite{pete}.

\begin{figure}
\centerline{\hbox{\psfig{figure=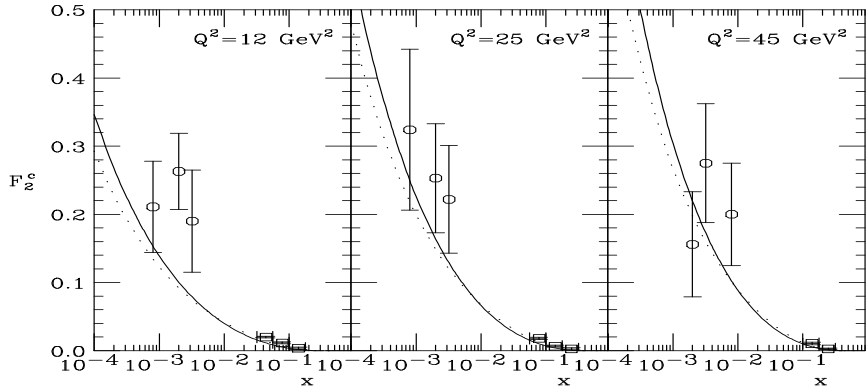,width=4.5in,height=2in}}}
\caption{Charm structure function $F_2(x,Q^2,m_c^2)$ as a function 
of $x$ for several $Q^2$ values.  Data is from DIS $D^o$ and $D^*$ 
production \protect\cite{h1}.  Theory is NLO QCD as described in text.}
\end{figure}
\section{Results}
Shown in Figure 1 is the charm contribution $F_2(x,Q^2,m_c^2)$ to the 
proton structure function as derived from inclusive $D^{*\pm}(2010)$ and 
$D^{o}(1864)$ production by H1 \cite{h1} (circles) as a function of 
$x$ at several $Q^2$ values.  
The EMC \cite{emc} data are also shown (boxes).  
The solid lines are NLO QCD \cite{hs} using GRV \cite{grv} parton distribution 
functions.  The dotted lines are NLO QCD using CTEQ \cite{cteq}.  
In both cases $\mu=\mu_f=\mu_r=\sqrt{Q^2+4m_c^2}$.  
The sensitivity of the NLO calculation with respect to variations in 
the renormalization (factorization) scale is greatly reduced relative 
to the LO calculation amounting to $\pm 10\%$ at small $x$ when the 
scale is varied between $2m_c$ and $2\sqrt{Q^2+4m_c^2}$.  
At small $Q^2$ the largest uncertainty is from the charm quark mass.  
A $\pm 10\%$ variation about the nominal value of $1.5\, {\rm GeV}$ used here 
results in a variation of $\pm 15\%$ at small $x$ and $Q^2=12\, {\rm GeV}$ 
which decreases to $\pm 8\%$ at $Q^2=45\, {\rm GeV}$.

In Figure 2 $D^{*}$ production cross sections for the
kinematic range $5\, {\rm GeV}^2<Q^2<100\, {\rm GeV}^2$ and $0<y<0.7$ with
$1.3\, {\rm GeV} < p_t^{D^{*}} < 9\, {\rm GeV}$ and $|\eta^{D^{*}}| < 1.5$ 
are shown.  
The data, shown as a function of $(a)\, p_t^{D^{*}}$, $(b)\, \eta^{D^{*}}$,
$(c)\, W$, and $(d)\, Q^2$, are from a recent ZEUS analysis \cite{zeus}.
The various lines are NLO QCD \cite{hs} using GRV \cite{grv} distributions 
supplemented with Peterson fragmentation \cite{pete}.  
The dashed line is for $\mu=2m_c$, $m_c=1.35\,{\rm GeV}$ and $\epsilon=0.035$ 
and the solid line is $\mu=2\sqrt{Q^2+4m_c^2}$, $m_c=1.65\,{\rm GeV}$ and 
$\epsilon=0.06$.  The agreement between the data and the theory is quite 
reasonable at this early stage of analysis.

\begin{figure}
\centerline{\hbox{\psfig{figure=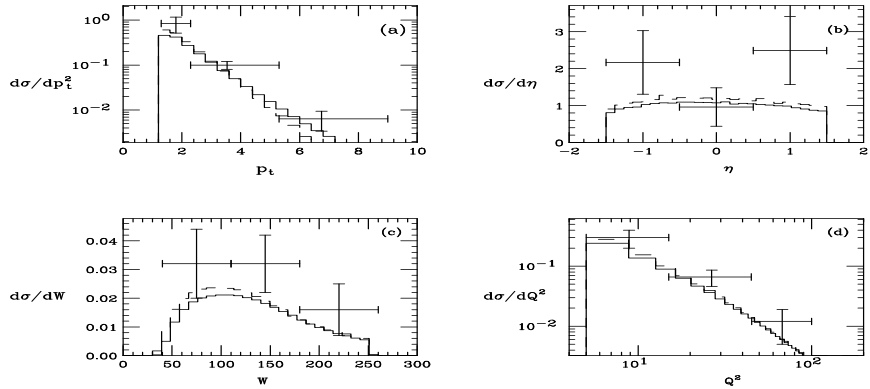,width=4.5in,height=2in}}}
\caption{Differential cross sections compared to ZEUS \protect\cite{zeus} 
DIS $D^*$ data.}
\end{figure}
\section{Conclusion}
The differential distributions presented here required placing cuts on 
several different variables.  The flexibility of the Monte Carlo approach has 
enabled us to calculate all of these simultaneously.  With the analysis of 
the 1995 data still to come and the accumulation of more data in the future 
we look forward to incorporating charm production in deeply inelastic 
scattering into a global fit to constrain the gluon distribution at low $x$.

\section*{Acknowledgments}I thank Juan Pablo Fernandez for discussions 
concerning the ZEUS data.  I also thank the Theory Group at Fermilab for 
a productive visit this summer.

\end{document}